\documentclass[11pt]{article}
\usepackage{amsmath,amssymb,graphicx,bm}
\newcommand{\be}{\begin{equation}}
\newcommand{\ee}{\end{equation}}

\begin{document}

\title{Asymptotics of Selberg-like integrals by lattice path counting}
\author{Marcel Novaes\\ {\small Departamento de F\'isica, Universidade Federal de S\~ao
Carlos, S\~ao Carlos, SP, 13565-905, Brazil.}}

\maketitle

\begin{abstract}
We obtain explicit expressions for positive integer moments of the
probability density of eigenvalues of the Jacobi and Laguerre random
matrix ensembles, in the asymptotic regime of large dimension. These
densities are closely related to the Selberg and Selberg-like
multidimensional integrals. Our method of solution is combinatorial:
it consists in the enumeration of certain classes of lattice paths
associated to the solution of recurrence relations.
\end{abstract}

\section{Introduction}

Let $N$ be a positive integer and $\mathcal{C}=[0,1]^N$ be an
$N$-dimensional hypercube. The integral \be
S(\alpha,\gamma,\beta)=\int_{\mathcal{C}}
|\Delta(T)|^\beta\prod_{i=1}^N
T_i^{\alpha-1}(1-T_i)^{\gamma-1}dT_i,\ee where
$\Delta(T)=\prod_{1\le i<j\le N}(T_i-T_j)$ is the Vandermonde
determinant, was evaluated by Selberg. It has found applications in
many different areas of mathematics \cite{dyson,kaneko,keating} and
physics \cite{hall,beenakker,bose}, in particular in the theory of
random matrices \cite{mehta,forrester}. A recent review of
applications as well as of the history of the field can be found in
\cite{forr}. Within the context of random matrix theory the constant
$\beta$ identifies the orthogonal, unitary and symplectic
universality classes and takes value in the set $\{1,2,4\}$. The
normalized function \be\label{density} P^{\alpha,\gamma}_{\beta}(T)=
\frac{1}{S(\alpha,\gamma,\beta)}|\Delta(T)|^\beta
\prod_{i=1}^NT_i^{\alpha-1}(1-T_i)^{\gamma-1}\ee is the joint
probability density of the eigenvalues of matrices from the Jacobi
$\beta$-Ensemble \cite{forrester}. The average value of any function
of the eigenvalues $f(T)$ is given by the multiple integral \be
\langle f(T)\rangle = \int_{\mathcal{C}}f(T)
P^{\alpha,\gamma}_{\beta}(T)d^NT,\ee with $d^NT=\prod_i dT_i$.

One particular application involves quantum electronic transport in
mesoscopic structures \cite{beenakker}. If there are $N_1$ incoming
channels and $N_2$ outgoing channels, the system may be described by
a $N_2\times N_1$ transmission matrix $t$ whose element $t_{ij}$ is
the probability amplitude of transmission from channel $j$ to
channel $i$. For systems with chaotic classical dynamics the random
matrix approach to the problem consists in assuming the system's
unitary $S$-matrix to be a random element of a circular
$\beta$-ensemble. This is equivalent to assuming the hermitian
matrix $\mathcal{T}=tt^\dag$ to be uniformly distributed in the
Jacobi $\beta$-Ensemble \cite{forr2006} with $N={\rm
min}\{N_1,N_2\}$, $\alpha=\frac{\beta}{2}(|N_2-N_1|+1)$ and
$\gamma=1$.

In the above context (with $\gamma=1$) the average value of quantities like \be\label{traces} {\rm Tr}[\mathcal{T}^n]=\sum_{i=1}^N T_i^n\ee can
be used to characterize universal statistical properties of quantum
chaotic transport. Exact results appeared for small $n$ in
\cite{small,shot,prb75mn2007,savin} and for general $n$ in \cite{vivo,prb78mn2008}. In this last
work the present author used a result of Kaneko \cite{kaneko} and
Kadell \cite{kadell} which gives the average value of any Schur
function of the eigenvalues, an approach that was later
taken further in \cite{savin2,luque}. For large numbers of channels,
a generating function for the average value of (\ref{traces}) was
presented in \cite{brouwer}, and an explicit expression appeared in
\cite{prb75mn2007}.

We are concerned with the asymptotic regime $N\gg 1$ of the average
value of (\ref{traces}) with respect to the probability density
(\ref{density}). Special cases of this problem were recently
discussed in \cite{luque2}, by using the method from
\cite{prb78mn2008} and taking $N\gg 1$ in the last step. It was then
noticed that the results had combinatorial interpretations, but no
reason for that was provided. The authors of \cite{luque2} also
conjectured that in this regime the average has a factorization
property, i.e. $\langle f(T)g(T)\rangle$ becomes asymptotically
equal to the product $\langle f(T)\rangle\langle g(T)\rangle$ if the
functions $f(T)$ and $g(T)$ depend on disjoint sets of variables. If
this is true, then the quantity $\langle T_1^n\rangle$ becomes
indeed the most interesting one to compute.

We also consider the asymptotic regime $N\gg 1$ of the average value
of (\ref{traces}) within the Laguerre ensemble of random matrices.
It that case the eigenvalues belong to $[0,\infty)$ and have a joint
probability density given by \be\label{densLag}
L^{\alpha,\epsilon}_{\beta}(T)=\frac{1}{Y(\alpha,\beta,\epsilon)}
|\Delta(T)|^\beta\prod_{i=1}^N T_i^{\alpha}e^{- \epsilon T_i}, \ee
where $Y(\alpha,\beta,\epsilon)$ is a known normalization constant
which is given by a Selberg-like integral. This also finds an
application in the area of quantum chaotic transport
\cite{timesprl,times}. The eigenvalues of the Wigner-Smith
time-delay matrix $Q=-i\hbar S^{-1}\partial S/\partial E$ (where $E$
is the energy) are called proper delay times, $T_i$. The inverse
delay times $1/T_i$ are distributed according to the Laguerre
ensemble with $\alpha=\beta N/2$ and $\epsilon=-\beta N/(2\gamma)$
where $\gamma$ is the classical decay rate of the system.

\section{Statement of results}

We obtain an explicit formula for $\langle T_1^n\rangle$, in the
asymptotic regime, valid for arbitrary values of $\beta$ (not
restricted to the set $\{1,2,4\}$) and for arbitrary parameters
$\alpha$ and $\gamma$, which are allowed to grow linearly with $N$
as is sometimes required. Let $[x]$ denote the integer part of $x$
and let \be C_n={2n \choose n}\frac{1}{n+1}\ee be the Catalan
numbers. We show that, asymptotically, \be \label{jacobi} \langle
T_1^n\rangle =A_2\sum_{m=0}^{n-1}{n-1 \choose m}(-1)^m
A_3^{n-1-m}\sum_{k=0}^{\left[ \frac{m+1}{2}\right]}{m-k \choose
k}C_{m-k}(A_1A_2)^{m-k} (1-A_3)^k,\ee where \be A_1=\frac{\beta
N}{2(\alpha+\gamma+\beta N)}, \quad A_2=\frac{2\alpha+\beta
N}{2(\alpha+\gamma+\beta N)},\quad A_3=A_1+A_2.\ee

The above formula generalizes all special cases that have so far
been considered. Our derivation is essentially combinatorial, and
consists of finding a recurrence relation and then turning its
solution into the problem of enumerating certain lattice paths. In
the course of the calculation we prove the factorization conjecture
already mentioned, for polynomial functions $f(T)$ and $g(T)$.

For the Laguerre $\beta$-Ensemble (\ref{densLag}) we show that the
factorization conjecture holds as well and that the relevant
asymptotic average value is given by \be \langle T_1^n\rangle=
A_2\sum_{m=0}^{\left[\frac{n-1}{2}\right]}{n-1 \choose
2m}C_m(A_1A_2)^mA_3^{n-1-2m}\ee where \be A_1=\frac{\beta N}{2},
\quad A_2=\alpha+A_1,\quad A_3=A_1+A_2.\ee We also consider the
analogous problem for the distribution of proper delay times. It was
shown in \cite{gregjack}, by other means, that the solution contains
the Schr\"oder numbers; we present a combinatorial proof of that.

We note that the Laguerre and Jacobi ensembles are also widely
studied in the context of multivariate statistical analysis
\cite{forrester,muirhead}, where they are known as the Wishart
distribution and the multivariate beta distribution. Moments of the
latter were considered in \cite{stat}.

The paper is organized as follows. In Section 3 we consider the
quantum transport case, which involves Jacobi Ensembles with
$\gamma=1$. This is the simplest version of the problem. The lattice
paths involved are Dyck paths, which have only two types of steps.
In Section 4 we turn to the Laguerre Ensembles, which are second in
the complexity scale. The lattice paths involved are now Motzkin
paths, which have three types of steps. In Section 5 we consider the
distribution of proper delay times in quantum chaotic scattering
(closely related to Laguerre). The lattice paths that appear in that
case are Schr\"oder paths. Finally, in Section 6 we come to Jacobi
Ensembles with arbitrary $\gamma$, which involve lattice paths with
four types of steps.

\section{Jacobi Ensembles with $\gamma=1$}

Let $f(T)=\prod_{i=1}^{m-1} T_i^{\lambda_i}$. We will show that
$\langle f(T)T_m^n\rangle\approx\langle f(T)\rangle\langle T_m^n\rangle$
in the asymptotic regime $N\gg 1$. To that end let us define
$Q=f(T)T_m^{n+1}P^{\alpha}_\beta(T)$, with $n>\lambda_i$. Taking
into account that \be
\frac{dP^{\alpha}_\beta}{dT_m}=\frac{(\alpha-1)}{T_m}P^{\alpha}_\beta(T)+\beta
P^{\alpha}_\beta(T) \sum_{i\ne m}\frac{1}{T_m-T_i}\ee we can write
the derivative of $Q$ with respect to $T_m$, \be\label{dQ}
\frac{dQ}{dT_m}=f(T)P^{\alpha}_\beta(T)\left[(n+\alpha)T_m^{n}+\beta
\sum_{i\ne m}\frac{T_m^{n+1}}{T_m-T_i}\right].\ee Notice that \be
\delta Q=\int_0^1 dT_m \frac{dQ}{dT_m}\ee does not depend on the
value of $n$. We will use this fact to find a recurrence relation.
This idea goes back to Aomoto \cite{aomoto}.

The integral of (\ref{dQ}) is \be \int_\mathcal{C}
\frac{dQ}{dT_m}d^NT=(n+\alpha)\langle
f(T)T_m^n\rangle+\beta\sum_{i\ne m}\left\langle
\frac{f(T)T_m^{n+1}}{T_m-T_i}\right\rangle.\ee But
\be\label{crucial} \sum_{i\ne m}\left\langle
\frac{f(T)T_m^{n+1}}{T_m-T_i}\right\rangle=\sum_{i< m}\left\langle
\frac{f(T)T_m^{n+1}}{T_m-T_i}\right\rangle+(N-m)\left\langle
\frac{f(T)T_m^{n+1}}{T_m-T_{m+1}}\right\rangle.\ee By
symmetry,\begin{align}\label{RRn} \left\langle
\frac{f(T)T_m^{n+1}}{T_m-T_{m+1}}\right\rangle&=\frac{1}{2}\left\langle
f(T)\frac{T_m^{n+1}-T_{m+1}^{n+1}}{T_m-T_{m+1}}\right\rangle\\&=\frac{1}{2}\sum_{a=0}^{n}
\big\langle f(T) T_m^{n-a}T_{m+1}^{a}\big\rangle. \end{align} Also,
in the same vein \be \left\langle
\frac{f(T)T_m^{n+1}}{T_m-T_i}\right\rangle=\frac{1}{2}\sum_{a=0}^{n-\lambda_i}
\big\langle f(T) T_i^{a}T_m^{n-a}\big\rangle.\ee Every term in
(\ref{crucial}) is of the same order in the limit $N\to\infty$, and
we can thus make the approximation \be \sum_{i\ne
m}\left\langle\frac{ f(T)T_m^{n+1}}{T_m-T_i}\right\rangle=
N\left\langle
\frac{f(T)T_m^{n+1}}{T_m-T_{m+1}}\right\rangle+O(1).\ee This is the
crucial step towards the factorization property, because we have
just ignored the terms of the sum containing the variables that
appear in $f(T)$. They are the ones that obstruct the factorization
for finite $N$.

From now on we simply ignore the error terms and keep in mind that
the limit $N\to\infty$ is implicit. Defining the function \be
D^{i,j}_{n}=\sum_{a=1}^{n-1}T_i^{n-a}T_j^a\ee we have \be\label{eq1}
\int_\mathcal{C} \frac{dQ}{dT_m}d^NT=\left[\alpha+\beta
N\left(1-\frac{\delta_{n,0}}{2}\right)\right]\langle
f(T)T_m^n\rangle+\frac{\beta N}{2}\langle f(T)
D^{m,m+1}_{n}\rangle,\ee where we have neglected $n$ against
$\alpha+\beta N$ (remember that $\alpha$ may grow with $N$).

We have seen that the above does not depend on the value of $n$. In
particular, for $n=0$ it gives $(\alpha+\beta N/2)\langle
f(T)\rangle$ and for $n=1$ it gives $(\alpha+\beta N)\langle
f(T)T_m\rangle.$ Comparing the formulas for general $n$ and $n=0$ we
obtain \be\label{recur} \langle f(T)T_m^n\rangle=A_2\langle
f(T)\rangle -A_1\langle f(T)D_{n}^{m,m+1}\rangle,\ee where \be
A_1=\frac{\beta N}{2(\alpha+\beta N)},\quad A_2=\frac{2\alpha+\beta
N}{2(\alpha+\beta N)}.\ee Recurrence relation (\ref{recur}) proves
the factorization conjecture, at least for polynomials. This is
because the function $f(T)$ does not change when the relation is
iterated, so it should be clear that it results in $\langle
f(T)T_m^n\rangle=\langle f(T)\rangle\langle T_m^n\rangle$.

The use of letters $A_1,A_2$ comes from the quantum transport
setting, where if $N\gg 1$ we obtain $A_i\approx N_i/(N_1+N_2)$.
Also, in that case $\alpha$ is proportional to $\beta$, causing the
final result to be independent of $\beta$. Notice that if $\alpha$
is held fixed in the limit of large $N$ we end up with
$A_1=A_2=1/2$.

We now turn to our main objective, which is the calculation of the
basic quantity $\langle T^n\rangle$. We omit the index of the
variable since it is irrelevant. Let us define \be\label{Rn}
\mathcal{D}_{n}=\sum_{a=1}^{n-1}\langle T^{n-a}\rangle\langle
T^a\rangle.\ee We have $\langle T^0\rangle=1$, $\langle
T^1\rangle=A_2$ and the recurrence relation \be\label{recur1}
\langle T^n\rangle=A_2-A_1\mathcal{D}_{n},\quad n\ge 2.\ee We may
write this as \be \langle T^n\rangle=A_2-A_1M_n+2A_1\langle
T^{n}\rangle,\quad n\ge 1,\ee where we have defined \be
M_n=\sum_{a=0}^{n}\langle T^{n-a}\rangle\langle T^a\rangle.\ee

It is possible to obtain the ordinary generating function  \be
F(x)=1+\sum_{n\ge 1}\langle T^n\rangle x^n.\ee Since \be
[F(x)]^2=1+\sum_{n\ge 1} M_n x^n\ee we get the algebraic relation
\be F=1+\frac{A_2 x}{1-x}+2A_1(F-1)-A_1(F^2-1),\ee which can be
solved to give \be F(x)=1-\frac{1}{2A_1}+
\frac{1}{2A_1}\sqrt{1+\frac{4A_1A_2x}{1-x}}.\ee

The generating function of course provides $\langle T^n\rangle$ for
any value of $n$. However, an explicit formula for this quantity can
be obtained by representing the recurrence relation pictorially and
turning its solution into a lattice path counting problem.

Let us start with a brief example. The recurrence relation
(\ref{recur1}) gives \be\langle T_1^4\rangle=A_2-A_1(\langle
T_1^3T_2\rangle+\langle T_1^2T_2^2\rangle+\langle
T_1T_2^3\rangle).\ee The coefficient of $A_1$ contains all ordered
partitions of $4$ into two parts. In the next step we have \be
A_2-A_1\Big[A_2(\langle T_1\rangle+\langle T_1^2\rangle+\langle
T_1^3\rangle)-A_1(\langle T_1T_2T_3^2\rangle+\langle
T_1T_2^2T_3\rangle+\langle T_1^2T_2T_3\rangle)\Big].\ee Now the
coefficient of $A_1^2$ has the ordered partitions of $4$ into three
parts. The coefficient of $A_1A_2$ has the ordered partitions --into
one part-- of the numbers up to three. At every step, when the power
of $A_1$ is increased the number of parts in the ordered partitions
also increases; on the other hand, when the power of $A_2$ increases
one part of the previous partitions is removed.

Let $[N,m]$ denote the set of all ordered partitions of $N$ into $m$
parts. We can represent a general step in the iteration of the
recurrence relation as
\begin{align*} [&N,m] \xrightarrow{A_1}[N,m+1]\\
&\downarrow A_2  \hspace{1.8cm}\downarrow A_2 \\ \bigcup_{n=1}^{N-1}
[n,&m-1]\xrightarrow{A_1}\bigcup_{n=1}^{N-1}[n,m]
\end{align*} Notice the commutativity of the diagram.
We may therefore  think of $A_1$ and $A_2$ as directions in which we
can move as we proceed with the calculation. If we further simplify
notation by writing $(N,m)=\bigcup_{n=1}^{N}[n,m]$ then Figure 1
codifies the recurrence relation.

\begin{figure}[t!]
\center
\includegraphics[scale=0.9,clip]{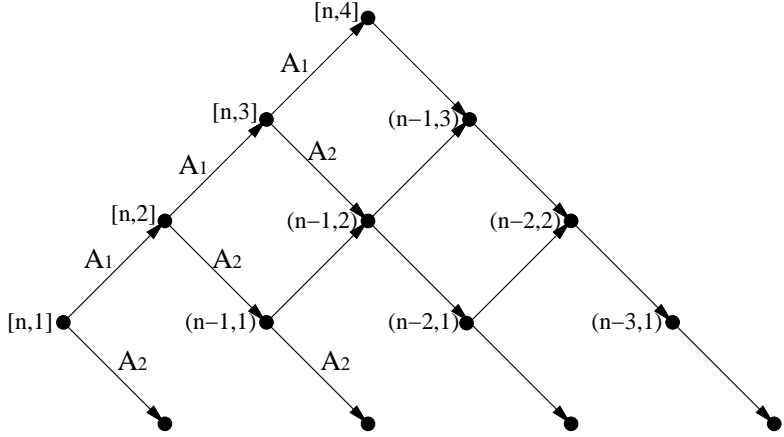}
\caption{The recurrence relation (\ref{recur1}) may be mapped into a
path-counting problem if we associate with $A_1$ and $A_2$ different
directions of movement. In the $A_1$ direction we increase the
number of parts of the ordered partitions, while in the $A_2$
direction we eliminate one part. If a path hits the lowest
horizontal line it stops. The number of terms is constant along
falling steps.}
\end{figure}

We start at $[n,1]$, which just represents $\langle T_1^n\rangle$.
Iteration of the recurrence relation then produces all possible
sequences of rising and falling steps that remain above the
$(\cdot,1)$ horizontal line. A path ends if and only if it drops
below that line. The final result for $\langle T^n\rangle$ will
therefore be equal to $A_2$ times a polynomial in $(-A_1A_2)$ of
degree $n-1$. The coefficient of $(-A_1A_2)^p$ contains two
contributions. First, the number of terms in the set $[n,p+1]$
(because moving in the $A_2$ direction does not change the number of
terms). This is the number of ordered partitions of $n$ into $p+1$
parts and is given by $\binom{n-1}{p}$. Second, the number of
different paths leading from $[n,1]$ to $(n-p,1)$. Clearly, the
relevant paths are Dyck paths: lattice paths with steps $(1,1)$ and
$(1,-1)$ that never fall below the $x$ axis. The number of such
paths containing $2p$ steps is well known to be the Catalan number,
$C_p$. In conclusion, \be\label{average} \langle
T^n\rangle=A_2\sum_{p=0}^{n-1}\binom{n-1}{p}(-1)^pC_p(A_1A_2)^p.\ee
This calculation has been sketched previously in \cite{prb75mn2007}.

\section{Laguerre Ensembles}

Before tackling the general Jacobi ensembles, we first consider the
Laguerre case. We now have the kernel \be
L^{\alpha,\epsilon}_\beta(T)=\frac{1}{Y}|\Delta|^\beta \prod_{i=1}^N
T_i^{\alpha}e^{-\epsilon T_i},\ee whose derivative is \be
\frac{dL^{\alpha,\epsilon}_\beta}{dT_m}=\frac{\alpha}{T_m}
L^{\alpha,\epsilon}_\beta(T)+\beta L^{\alpha,\epsilon}_\beta(T)
\sum_{i\ne m}\frac{1}{T_m-T_i}-\epsilon
L^{\alpha,\epsilon}_\beta(T).\ee The integration domain is now
$\mathcal{C}=[0,\infty)^N$. If we define
$Q=f(T)T_m^{n+1}L^{\alpha,\epsilon}_\beta(T)$ it is easy to see that
the value of $\int_{0}^\infty\frac{dQ}{dT_m}dT_m$ is always zero,
for any value of $n$. Proceeding analogously to what we did to
arrive at Eq.(\ref{eq1}), we find that \begin{align}\label{lag}
\int_{\mathcal{C}}\frac{dQ}{dT_m}d^NT=&\left[\alpha+\beta
N\left(1-\frac{\delta_{n,0}}{2}\right)\right]\langle
f(T)T_m^n\rangle\nonumber\\&+\frac{\beta N}{2}\langle f(T)
D^{m,m+1}_{n}\rangle-\epsilon\left\langle
f(T)T_m^{n+1}\right\rangle=0.\end{align}

The above equation implies that the factorization conjecture holds
again, since $f(T)$ is not affected by the iteration. Let us define
the new variables \be A_1=\frac{\beta N}{2\epsilon},\quad
A_2=\frac{\alpha}{\epsilon}+A_1,\quad A_3=A_1+A_2.\ee Eq.
(\ref{lag}) gives in particular $\langle T\rangle=A_2$ and $\langle
T^2\rangle=A_3\langle T\rangle$. For general $n$ we have \be\langle
T^n\rangle=A_3\langle T^{n-1}\rangle+A_1\mathcal{D}_{n-1},\ee where
$\mathcal{D}_{n}$ has been defined in Eq.(\ref{Rn}). Telescoping
this equation we obtain the recurrence relation \be\label{recur3}
\langle T^n\rangle =A_2A_3^{n-1}+A_1\sum_{k=0}^{n-2}
A_3^k\mathcal{D}_{n-k-1}.\ee  In terms of the generating function
$F(x)=\sum_{n\ge 0}\langle T^n\rangle x^n$ it is elementary to
derive the simple quadratic algebraic relation \be F=1+\frac{A_2
x+A_1x(F-1)^2}{1-A_3 x}.\ee

The recurrence relation (\ref{recur3}) can also be interpreted in
terms of lattice paths. Again the constants $A_1$, $A_2$ and $A_3$
are associated with different directions of movement in the plane,
which take us from one set of ordered partitions to another. In the
$A_2$ direction we simply remove one of the parts, as in the
previous section. In the $A_1$ direction we increase the number of
parts by one and decrease the number being partitioned by one. In
the $A_3$ direction we keep the number of parts constant and
decrease the number being partitioned by one. If we denote, as in
the previous section, by $[n,m]$ the set of all ordered partitions
of $n$ into $m$ parts, and define the union
$(N,m)=\bigcup_{n=1}^N[n,m]$, then the iteration of (\ref{recur3})
is depicted in Figure 2.

\begin{figure}[h!]
\center
\includegraphics[scale=0.65,clip]{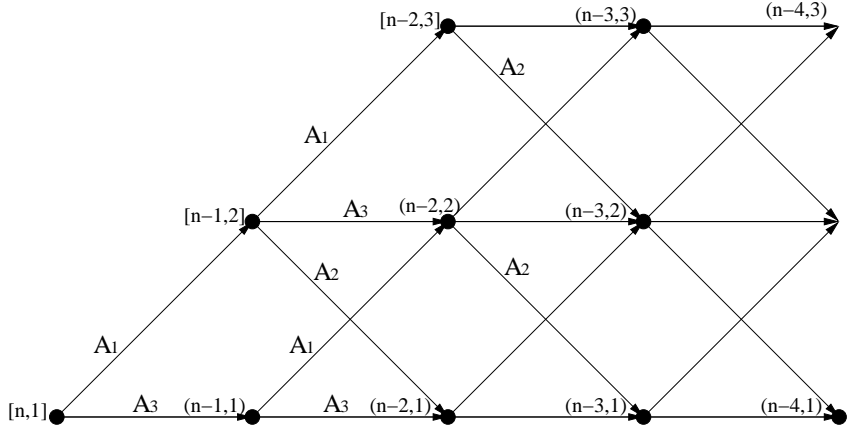}
\caption{The recurrence relation for the Laguerre ensemble
(\ref{recur3}) may be mapped into a path-counting problem if we
associate with $A_1$, $A_2$ and $A_3$ the rising, falling and
horizontal directions of movement, respectively. Notice that the
action of $A_1$ is different than in the previous section and in
Figure 1.}
\end{figure}

As a brief example, consider \begin{align}\langle
T_1^5\rangle=A_2A_3^4+A_1&\big(\langle T_1^3T_2\rangle+\langle
T_1^2T_2^2\rangle+\langle T_1T_2^3\rangle\nonumber\\&+A_3\langle
T_1^2T_2\rangle+A_3\langle T_1T_2^2\rangle+A_3^2\langle
T_1T_2\rangle\big).\end{align} We interpret this in the following
way: we can take two horizontal steps and one rising step to go to
the partitions of two into two parts, $A_1A_3^2\langle
T_1T_2\rangle$; we can take one horizontal step and one rising step
to go to the partitions of three into two parts, $A_1A_3(\langle
T_1^2T_2\rangle+\langle T_1T_2^2\rangle)$; we can take only one
rising step to reach partitions of four into two parts; finally, we
can take four horizontal steps and one falling step. In general,
from any given point we may go up or down, but before we do so we
can take any number of horizontal steps.

With one more iteration we find \begin{align} \langle
T_1^5\rangle=A_2A_3^4+&A_2A_1\langle T_1^3\rangle+A_2A_3A_1\langle
T_1^2\rangle+A_2A_3^2A_1\langle T_1\rangle+A_1^2\langle
T_1T_2T_3\rangle\nonumber\\&+A_2A_1A_3\langle
T_1^2\rangle+A_2A_3A_1A_3\langle T_1\rangle+A_2A_1A_3^2\langle
T_1\rangle. \end{align} It is instructive to treat the variables as
non-commutative because that makes it easier to visualize the paths.
So even though $A_2A_3^2A_1$, $A_2A_3A_1A_3$ and $A_2A_1A_3^2$ are
all equal, the lattice paths to which they correspond are different
(notice that the steps should be read from right to left).

The general structure of the calculation of $\langle T^n\rangle$ is
as follows. The last step is always $A_2$. The rest of the path
consists in $n-1$ steps, which clearly result in Motzkin paths:
lattice paths with steps $(1,1)$, $(1,-1)$ and $(1,0)$ that never
fall below the $x$ axis. The number of Motzkin paths of length $n$
containing exactly $m$ rising steps is given by \be
M_{n,m}=\binom{n}{2m}C_m,\ee where $C_m$ are the Catalan numbers. In
conclusion, \be \langle T^n\rangle= A_2
\sum_{m=0}^{\left[\frac{n-1}{2}\right]}M_{n-1,m}(A_1A_2)^mA_3^{n-1-2m}.\ee
Notice that if $\lim_{N\to\infty} \frac{\alpha}{N}=0$ we have
$\langle T^n\rangle=C_n \langle T\rangle^n$.

\section{Proper delay times}

It was shown in \cite{timesprl,times} that the proper delay times
$T_i$ associated with quantum scattering by a chaotic cavity are
distributed according to \be Z^{\alpha,\epsilon}_\beta(T)=
\frac{1}{W(\alpha,\beta,\epsilon)}|\Delta|^\beta \prod_{i=1}^N
T_i^{\alpha}e^{-\epsilon/T_i},\ee with $\alpha=-3\beta N/2+\beta-2$
and $\epsilon=-\beta N \tau_D/2$, where $\tau_D$ is the cavity's
classical dwell time and $N$ is the number of decay channels. The
average value of $T_1^n$ was computed in \cite{gregjack} by
integration against the displaced semicircle eigenvalue density, and
was found to be related to the large Schr\"oder numbers.

The calculation in this case is quite similar to that in the
previous Section. The derivative of the kernel is \be
\frac{dZ^{\alpha,\epsilon}_\beta}{dT_m}=\left[\frac{\alpha}{T_m}
+\beta \sum_{i\ne m}\frac{1}{T_m-T_i}+\frac{\epsilon}{T_m^2}
\right]Z^{\alpha,\epsilon}_\beta(T).\ee Defining
$Q=f(T)T_m^{n+1}Z^{\alpha,\epsilon}_\beta(T)$, the value of
$\int_{0}^\infty\frac{dQ}{dT_m}dT_m$ is again zero for any value of
$n$. Now we take $n>0$ and get \be\label{tim} (\alpha+\beta
N)\langle f(T)T_m^n\rangle+\frac{\beta N}{2}\langle f(T)
D^{m,m+1}_{n}\rangle+\epsilon\left\langle
f(T)T_m^{n-1}\right\rangle=0.\ee

The above equation implies that the factorization conjecture holds
again, since $f(T)$ is not affected by the iteration. Defining the
new variables \be A_1=-\frac{\beta N}{2(\alpha+\beta N)}= 1,\quad
A_2=A_3=-\frac{\epsilon}{\alpha+\beta N}= \tau_D,\ee Eq. (\ref{tim})
gives $\langle T\rangle=A_2$ and, in general, $\langle
T^n\rangle=A_3\langle T^{n-1}\rangle+A_1\mathcal{D}_{n}$.
Telescoping gives \be\label{recur4} \langle T^n\rangle
=A_2A_3^{n-1}+A_1\sum_{k=0}^{n-2} A_3^k\mathcal{D}_{n-k}.\ee

The recurrence relation (\ref{recur4}) is only slightly different
from (\ref{recur3}). The difference is that steps in the $A_1$
direction no longer change the number being partitioned. The
iteration of (\ref{recur4}) may be interpreted as in Figure 3: $A_1$
corresponds to a vertical step, while $A_2$ and $A_3$ correspond to
falling and horizontal steps, respectively. The paths involved in
the calculation of $\langle T^n\rangle$ are those going from $[n,1]$
to $[1,1]$ without falling below the initial horizontal level. Once
the step reaches $[1,1]$ a final $A_2$ step terminates it. The total
number of $A_2$ steps plus the total number of $A_3$ steps is always
equal to $n$. Since $A_1=1$ and $A_2=A_3=\tau_D$ the value of
$\langle T^n\rangle$ will be equal to $\tau_D^{n}$ times the number
of possible paths.

The solution to the enumeration problem posed above is nothing but
the (large) Schr\"oder number $R_n$. In order to see this, suppose
we turn every $A_1$ step from vertical to rising (i.e. from $(0,1)$
to $(1,1)$) and double every $A_3$ step (i.e. from $(1,0)$ to
$(2,0)$). The resulting path will always be a Schr\"oder path of
length $2n$. The map is bijective, so $R_n$ is the number we are
looking for. In conclusion, we have \be \langle T^n\rangle=R_n
\tau_D^n.\ee

\begin{figure}[t!]
\center
\includegraphics[scale=0.7,clip]{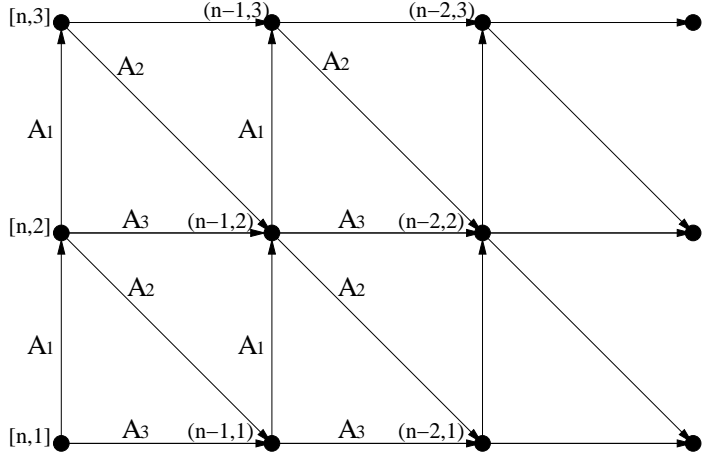}
\caption{The recurrence relation (\ref{recur4}) associated with
proper time delays is mapped into a path-counting problem by taking
$A_1$, $A_2$ and $A_3$ to be vertical, falling and horizontal types
of steps, respectively. These paths are in bijection to Schr\"oder
paths.}
\end{figure}

\section{Jacobi Ensembles}

We now come back to the general Jacobi Ensembles, when the Selberg
kernel is \be P^{\alpha,\gamma}_\beta(T)= \frac{1}{S}|\Delta|^\beta
\prod_{i=1}^N T_i^{\alpha-1}(1-T_i)^{\gamma-1}\ee and the
integration domain is again $\mathcal{C}=[0,1]^N$. In this case we
have \be
\frac{dP^{\alpha,\gamma}_\beta}{dT_m}=\frac{(\alpha-1)}{T_m}
P^{\alpha,\gamma}_\beta(T)+\beta P^{\alpha,\gamma}_\beta(T)
\sum_{i\ne m}\frac{1}{T_m-T_i}-
(\gamma-1)\frac{P^{\alpha,\gamma}_\beta(T)}{1-T_m}\ee and the
integral of $dQ/dT_m$ with
$Q=f(T)T_m^{n+1}P^{\alpha,\gamma}_\beta(T)$ is given by
\begin{align} \int_\mathcal{C}
\frac{dQ}{dT_m}d^NT=&\left[\alpha+\beta
N\left(1-\frac{\delta_{n,o}}{2}\right)\right]\langle
f(T)T_m^n\rangle\nonumber\\\label{q}&+\frac{\beta N}{2}\langle f(T)
D^{m,m+1}_{n}\rangle-(\gamma-1)\left\langle
f(T)\frac{T_m^{n+1}}{1-T_m}\right\rangle.\end{align} It remains
independent of $n$. We have again neglected $n$ compared to
$\alpha+\beta N$, and we may also approximate $\gamma-1\approx
\gamma$ since what matters is $\lim_{N\to\infty}\gamma/N$.

Let us define \be A_1=\frac{\beta N}{2(\alpha+\beta N+\gamma)},\quad
A_2=\frac{2\alpha+\beta N}{2(\alpha+\beta N+\gamma)},\quad
A_3=A_1+A_2.\ee Comparing the values of (\ref{q}) at $n=1$ and $n=0$
we have $\langle f(T) T_m\rangle=A_2\langle f(T)\rangle.$ Comparing
in general $n$ and $n-1$ we have \be \langle f(T)T_m^n\rangle
=A_3\langle f(T)T_m^{n-1}\rangle+A_1\langle
f(T)[D_{n-1}^{m,m+1}-D_n^{m,m+1}]\rangle.\ee The factorization
property clearly continues to hold for general $\gamma$.

Let us now consider $\langle T^n\rangle$. We have $\langle T\rangle
=A_2$. For general $n$ we may telescope the previous equation to
obtain \be\label{recur2} \langle T^n\rangle=A_2A_3^{n-1}+(1-A_3)A_1
\sum_{k=0}^{n-2}A_3^{k} \mathcal{D}_{n-k-1}-A_1\mathcal{D}_n,\ee
where $\mathcal{D}_{n}$ has been defined in Eq.(\ref{Rn}). For the
generating function $F(x)=\sum_{n\ge 0}\langle T^n\rangle x^n$ this
leads again to a quadratic algebraic relation \be F=1+\frac{A_2
x}{1-A_3 x}-A_1(F-1)^2\frac{1-x}{1-A_3 x}. \ee

In order to turn the recurrence relation into a lattice path problem
it is convenient to write it as \be \langle
T^n\rangle=A_2A_3^{n-1}+A_4\sum_{k=0}^{n-2}A_3^{k}\mathcal{D}_{n-k-1}
-A_1\mathcal{D}_{n},\ee where $A_4=A_1(1-A_3)$ is treated as an
independent variable. Comparing this equation to the ones obtained
in the previous sections, we see that $A_1$ and $A_2$ have the same
role they had in Section 3, namely $A_2$ removes one part of every
partition and $A_1$ increases the number of parts by one. $A_3$ has
the same role it had in Section 4: it decreases the number being
partitioned by one. Finally, $A_4$ now does what $A_1$ did in
Section 4, it decreases the number being partitioned by one while
increasing the number of parts by one. The general structure is as
depicted in Figure 4.

\begin{figure}[b!]
\center
\includegraphics[scale=0.65,clip]{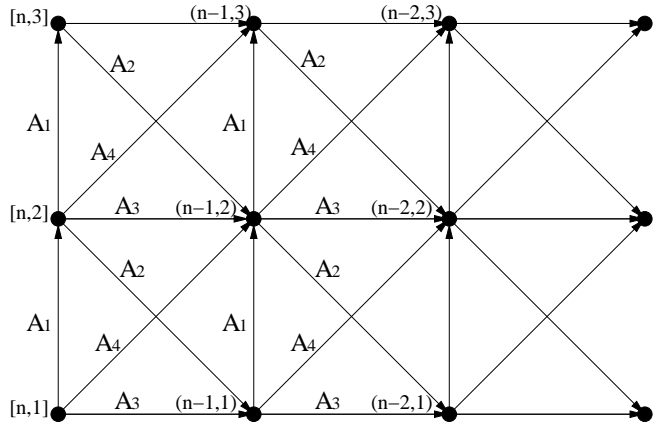}
\caption{The recurrence relation for the Jacobi ensemble corresponds
to lattice paths where $A_1$, $A_2$, $A_3$ and $A_4$ are
respectively associated with vertical, falling, horizontal and
rising steps.}
\end{figure}

We now have lattice paths with four possible steps: $(0,1)$,
$(1,-1)$, $(1,0)$ and $(1,1)$ corresponding respectively to $A_1$,
$A_2$, $A_3$ and $A_4$. The total displacement in the horizontal
direction must be $n-1$. Let us consider the situations in which
exactly $n-1-m$ of the horizontal moves come from $A_3$ steps. There
are ${n-1 \choose m}$ possibilities for assigning their position.
Let us suppose that exactly $k$ of the remaining steps are of type
$A_4$. We are thus reduced to counting the number of paths with $k$
raising steps, $m-k$ falling steps and $m-2k$ vertical steps. The
key to this enumeration is that every one of these paths may be
obtained from a Dyck path of length $2m-2k$, if we replace $m-2k$ of
its raising steps by vertical ones. The number of such Dyck paths is
$C_{m-k}$, and the number of possibilities to single out $m-2k$ of
the raising steps is ${m-k \choose k}$.

In conclusion, the final result is given by \be \langle
T^n\rangle=A_2 \sum_{m=0}^{n-1}{n-1 \choose m}
A_3^{n-1-m}\sum_{k=0}^{\left[ \frac{m+1}{2}\right]}{m-k \choose
k}C_{m-k}(-A_1A_2)^{m-2k}(A_4A_2)^k.\ee Replacing back $A_4$ by
$A_1(1-A_3)$ we arrive at the result (\ref{jacobi}). Naturally, when
$\lim_{N\to\infty}\gamma/N=0$ we have $A_3=1$ and this reduces to
(\ref{average}).


\begin{thebibliography}{99}

\bibitem{dyson} Dyson, F.J.: Statistical theory of energy levels
of complex systems I. J. Math. Phys. {\bf 3}, 140-–156 (1962).

\bibitem{kaneko} Kaneko, J.: Selberg integrals and hypergeometric
 functions associated with Jack polynomials. SIAM J. Math. Analysis, {\bf 24}, 1086--1110
(1993).

\bibitem{keating} Keating, J.P., Snaith, N.C.: Random matrix
 theory and $\zeta(1/2+it)$. Comm. Math. Phys. {\bf 214}, 57-–89 (2001).

\bibitem{hall} Di Francesco, P., Gaudin, M., Itzykson, C.,
 Lesage, F.: Laughlin's wave functions, Coulomb gases and expansions
  of the discriminant. Int. J. Mod. Phys. A {\bf 9}, 4257--4351 (1994).

\bibitem{beenakker} Beenakker, C.W.J.: Random matrix theory
 of quantum transport. Rev. Mod. Phys. {\bf 69}, 731--808 (1997).

\bibitem{bose} Forrester, P.J., Frankel, N.E., Garoni, T.M.:
 Random matrix averages and the impenetrable Bose gas in Dirichlet
  and Neumann boundary conditions. J. Math. Phys. {\bf 44}, 4157-4175 (2003).

\bibitem{mehta} Mehta, M.L.: {\it Random Matrices}. New York:
 Academic Press, 2004.

\bibitem{forrester} Forrester, P.J.: {\it Log-Gases and Random
 Matrices}. Princeton: Princeton University Press, 2010.

\bibitem{forr} Forrester, P.J., Warnaar, S.O.: The importance
 of the Selberg integral. Bull. Am. Math. Soc. {\bf 45}, 489--534 (2008).

\bibitem{forr2006} Forrester, P.J.: Quantum conductance problems
 and the Jacobi ensemble. J. Phys. A {\bf 39}, 6861--6870 (2006).

\bibitem{small} Baranger, H.U., Mello, P.A.: Mesoscopic transport
 through chaotic cavities: A random S-matrix theory approach. Phys. Rev. Lett. {\bf 73}, 142-–145 (1994).

\bibitem{shot} Savin, D.V., Sommers, H.-J.: Shot noise in chaotic
 cavities with an arbitrary number of open channels, Phys. Rev. B {\bf 73}, 081307 (2006).

\bibitem{prb75mn2007} Novaes, M.: Full counting statistics of chaotic
 cavities with many open channels. Phys. Rev. B {\bf 75}, 073304 (2007).

\bibitem{savin} Savin, D.V., Sommers, H.-J., Wieczorek, W.: Statistics
 of quantum transport in chaotic cavities with broken time-reversal symmetry. Phys. Rev. B {\bf 77}, 125332 (2008).

\bibitem{vivo} Vivo, P., Vivo, E.: Transmission eigenvalue densities
 and moments in chaotic cavities from random matrix theory. J. Phys. A {\bf 41}, 122004 (2008).

\bibitem{prb78mn2008} Novaes, M.: Statistics of quantum transport
 in chaotic cavities with broken time-reversal symmetry. Phys. Rev. B {\bf 78}, 035337 (2008).

\bibitem{kadell} Kadell, K.W.J.: The Selberg-Jack symmetric
 functions, Adv. Math. {\bf 130}, 33--102 (1997).

\bibitem{savin2} Khoruzhenko, B.A., Savin, D.V., Sommers, H.-J.:
 Systematic approach to statistics of conductance and shot-noise in chaotic cavities. Phys. Rev. B {\bf 80}, 125301 (2009).

\bibitem{luque} Luque, J.-G., Vivo, P.: Nonlinear random matrix
 statistics, symmetric functions and hyperdeterminants. J. Phys. A {\bf 43}, 085213 (2010).

\bibitem{brouwer} Brouwer, P.W., Beenakker, C.W.J.: Diagrammatic
 method of integration over the unitary group, with applications
  to quantum transport in mesoscopic systems. J. Math. Phys. {\bf 37}, 4904--4934 (1996).

\bibitem{luque2} Carr\'e, C., Deneufchatel, M., Luque, J.-G., Vivo,
P.: Asymptotics of Selberg-like integrals: The unitary case and
Newton's interpolation formula, arXiv:1003.5996v1.

\bibitem{timesprl} Brouwer, P.W., Frahm, K.M., Beenakker, C.W.J.:
Quantum Mechanical Time-Delay Matrix in Chaotic Scattering. Phys.
Rev. Lett. {\bf 78}, 4737-4740 (1997).

\bibitem{times} Brouwer, P.W., Frahm, K.M., Beenakker, C.W.J.:
Distribution of the quantum mechanical time-delay matrix for a
chaotic cavity. Waves Random Media {\bf 9}, 91--104 (1999).

\bibitem{gregjack} Berkolaiko, G., Kuipers, J.: Moments of delay
times. J. Phys. A {\bf 43}, 035101 (2010).

\bibitem{muirhead} Muirhead, R.J.: {\it Aspects of Multivariate
Statistical Theory}, Chapter 3. Hoboken, Wiley, 2005.

\bibitem{stat} Bai, Z.D, Yin, Z.Q., Krishnaiah, P.R.: On the
limiting empirical distribution function of the eigenvalues of a
multivariate F matrix. Theory Probab. Appl. {\bf 32}, 490--500
(1985).

\bibitem{aomoto} Aomoto, K.: Jacobi polynomials associated with
Selberg integrals. SIAM J. Math. Analysis, {\bf 18}, 545--549
(1987).

\end{thebibliography}
\end{document}